\documentclass[sigconf,review = false]{acmart}

\usepackage{soul}

\usepackage{subcaption}
\AtBeginDocument{%
	}

\setcopyright{acmcopyright}
\copyrightyear{2023}
\acmYear{2023}
\acmDOI{XXXXXXX.XXXXXXX}

\acmConference[IEEE]{Popularity Estimation and New Bundle Generation using Content and Context based Embeddings}
\acmPrice{15.00}
\acmISBN{978-1-4503-XXXX-X/18/06}

\begin{document}
	
	\title{Popularity Estimation and New Bundle Generation using Content and Context based Embeddings}
	
	
	\author{Dr. Ashutosh Nayak}
	\authornote{All authors contributed equally to this research.}
	\affiliation{%
		\institution{Samsung Research Institute Bangalore, SRIB}
		\country{India}
	}
	
	\author{Prajwal NJ}
	\affiliation{%
		\institution{RV College of Engineering}
		\city{Bangalore}
		\country{India}}
	
	\author{Sameeksha Keshav}
	\affiliation{%
		\institution{RV College of Engineering}
		\city{Bangalore}
		\country{India}}
	
	\author{Dr. Kavitha S.N.}
	\affiliation{%
		\institution{RV College of Engineering}
		\city{Bangalore}
		\country{India}}
	
	\author{Dr. Roja Reddy}
	\affiliation{%
		\institution{RV College of Engineering}
		\city{Bangalore}
		\country{India}}
	
	\author{Dr. Rajasekhara Reddy Duvvuru Muni}
	\authornotemark[1]
	\email{raja.duvvuru@samsung.com}
	\affiliation{%
		\institution{Samsung Research Institute Bangalore, SRIB}
		\country{India}
	}

	\renewcommand{\shortauthors}{Ashutosh et al.}
	
	\begin{abstract}
		Recommender systems create enormous value for businesses and their consumers. They increase revenue for businesses while improving the consumer experience by recommending relevant products amidst huge product base. Product bundling is an exciting development in the field of product recommendations. It aims at generating new bundles and recommending exciting and relevant bundles to their consumers. Unlike traditional recommender systems that recommend single items to consumers, product bundling aims at targeting a bundle, or a set of items, to the consumers. While bundle recommendation has attracted significant research interest recently, extant literature on bundle generation is scarce. Moreover, metrics to identify if a bundle is popular or not is not well studied. In this work, we aim to fulfill this gap by introducing new bundle popularity metrics based on sales, consumer experience and item diversity in a bundle. We use these metrics in the methodology proposed in this paper to generate new bundles for mobile games using content aware and context aware embeddings. We use open-source Steam Games dataset for our analysis. Our experiments indicate that we can generate new bundles that can outperform the existing bundles on the popularity metrics by 32\% - 44\%. Our experiments are computationally efficient and the proposed methodology is generic that can be extended to other bundling problems e.g. product bundling, music bundling.
	\end{abstract}

	\ccsdesc[500]{Computing Methods~Artificial Intelligence}
	\ccsdesc[300]{Search Methodologies~Heuristic function construction}
	
	\keywords{Bundle Generation, Popularity Metrics, Context Aware Sentence Embeddings, Content Aware Item Embeddingsy}
	
	\received{May 2024}
	\received[revised]{May 2024}
	\received[accepted]{May 2024}
	
	\maketitle
	
\section{Introduction}\label{sec:introduction}
Product Recommender Systems (RS) create huge business values for e-commerce businesses. RS aims at recommending relevant products to its consumers. This helps in increasing consumer satisfaction as consumers can find relevant products in shorter time leading to downstream benefits for the firms like increased sales. A new paradigm of product recommendation, called Bundle recommendation is attracting significant interest from industry and academia. A bundle is a set of items offered together as a purchase entity. Bundle recommendation involves recommending bundles to consumers as against recommending single items. Although bundle recommendation has been studied in the literature, the question of how to evaluate and create new bundles is an open area of research. Current literature on bundle generation is scarce and we aim to fill this gap in this research by proposing a novel holistic approach for evaluating bundles and generating new bundles.

\begin{figure}[!h]
	\centerline{\includegraphics[width=\linewidth]{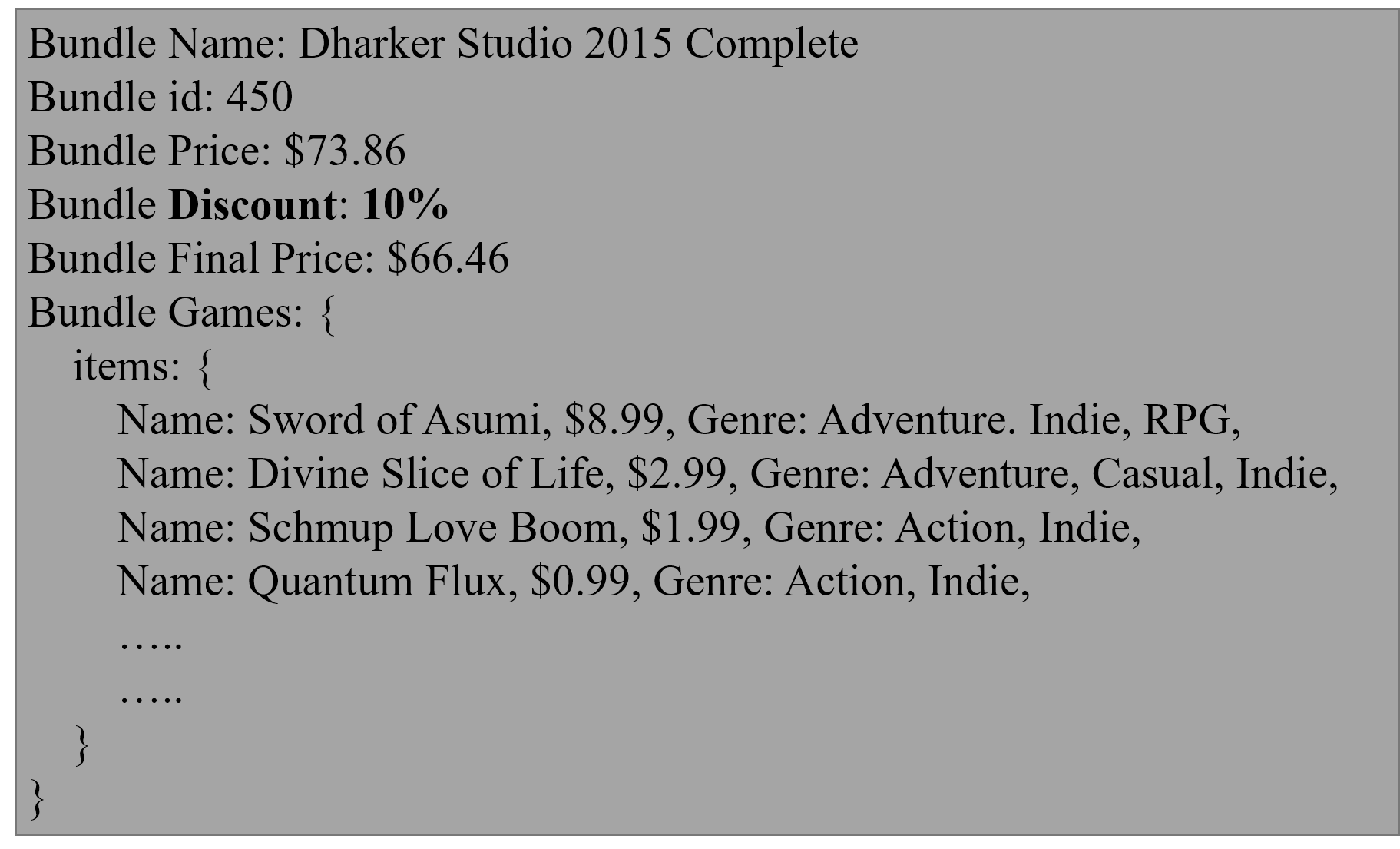}}
	\caption{Example of a Bundle in Steam Games}
	\label{fig:bundleExample}
\end{figure}

A bundles includes connected items grouped together to make it attractive to consumers for purchase in one go. There are two types of bundling - dynamic and static. Dynamic bundling recommends bundles based on current consumer session (e.g. Amazon). Static bundling suggests pre-constructed bundles (e.g. Best Buy). We focus on static bundling problem where managers decide how many and which bundles to include in their e-commerce business. Example of a bundle is shown in Figure~\ref{fig:bundleExample}. The example is taken from steam games where 3 games are grouped as bundle. Figure~\ref{fig:bundleExample} also shows the characteristic of a bundle, e.g. bundle name, bundle price and bundle discount. Bundles are beneficial for both the consumers and the firms offering bundles. Bundles are generally coupled with discount, thus consumers can buy the products at a discounted price~\cite{yadav1993buyers}. Furthermore, right bundles reduces consumers' search time. In e-commerce businesses, unit shipping cost reduces as the weight of the orders increase, thus multiple products purchase reduces the shipping cost for the firms. With digital product bundling, it reduces the marketing cost for firms as they can couple new products with an existing bundle, thus exposing new products to the consumers. 

Existing RS has enhanced our ability to recommend relevant bundles to consumers. Thereby, given a set of bundles, modern RS can provide good bundle recommendations to its consumers. But how these bundles are created is still an interesting research problem. Some common factors which might affect the quality of a bundle include number of items in a bundle, price of a bundle, discount offered or age of the items. The traditional method of creating bundles manually by managers is not generalizeable and not scalable in e-commerce businesses with thousands of products. The statistical approach of combining products in the same basket from past consumer purchases suffers from sparsity as the proportion of items purchased together as compared to the total number of items is very low. This method also lacks a common theme in a basket as a consumer may purchase unrelated items in a basket. () use a manual interpreters to understand the theme of a bundle. However, this approach could still suffer with new products as they use past purchase baskets to create bundles.

To further the research in creating new bundles agnostic to the domain, we consider two fundamental limitations of bundle generation. First, there is lack of a clear definition to identify the quality of a bundle. We define quality of a bundle as the likelihood of a bundle to be popular among its consumers. We use quality and popularity interchangeably in this paper. Extant literature considers clicks of a bundle among recommended bundles as a measure for bundle quality. However, it depends on the algorithm used for recommending the bundles to consumers. For example, if a bundle is never recommended to users, it will never be clicked, hence the implied value of a bundle is low. Thus it is critical to define a metric for calculating the quality of a given bundle. Second, there is lack of understanding of why a given set of items is clicked or not clicked. There are different types of bundles as items in a bundle can be -- complimentary items (toothbrush and toothpaste), similar items (funny movies), sequential items (flight and hotel booking) or items from same product family (Harry Potter book series). Thus, understanding the relation between items in a bundle is critical for sampling items for new bundle creation. 

In this paper, we use embeddings to sample items for bundle creation. Embeddings are vector representation of information about an item in a latent $D$-Dimensional embedding space such that similar items are spherically and directionally close to each other(). Embeddings can condense explicit and implicit information about an item in numerical data format. Explicit information about an item includes name, features, texts or visuals. Implicit information includes behavioral outlook of consumers e.g. products bought/viewed together by consumers.  In this paper, we use Steam Data~\cite{pathak2017generating} for our analysis and model development. We use different combinations of textual information to generate sentence embeddings~\cite{reimers2019sentence}. In this study, we analyse different embeddings based on explicit-only, implicit-only or explicit-implicit information. We use these embeddings to sample items from the complete item set to create new bundles using different strategies. To the best knowledge of the authors, this is the first work that uses embeddings to optimize product bundling

Overall, we aim to propose a holistic approach to creating new bundles. Towards this goal, this work builds different metric to measure the quality of bundles and use this capability to create new optimized bundles. We believe the methodology proposed in this paper for creating new bundles is generalizeable and our work is more important in sectors when the number of items is large and new items are added regularly, for example, creating a songs playlist (Spotify), movie marathon list (Netflix), technical articles (Medium) or e-commerce products (Amazon). This work also presents the first evaluation of different quality metrics for bundles. The major contribution of this paper are:
\begin{itemize}
	\item We propose new metrics for measuring the quality of a bundle and present corresponding advantages and disadvantages
	\item We present and evaluate the use of different embeddings for sampling items for creating new bundles
	\item We construct machine learning models to predict the quality of a bundle to enable optimizing a bundle for different popularity metrics
	\item We introduced a novel methodology to create popular bundles and validate the findings using different experiments. 
\end{itemize}

The rest of the paper is organized as: We discuss extant literature in Section~\ref{sec:relatedWork}. We lay out the detailed problem definition in Section~\ref{sec:problemFormulation}. We then present our proposed methodology on evaluating a bundle and generating new bundles in Section~\ref{sec:proposedMethodology}. Results from our analysis are shown in Section~\ref{sec:results}. We conclude the paper with discussion and scope for future research in Section~\ref{sec:conclusion}.

\section{Related Work}\label{sec:relatedWork}
In this Section, we discuss relevant extant works in the field of product bundling. Zhu \textit{et al.}~\cite{sun2022revisiting} presented a comprehensive work on product bundling problem. Product bundling is a multi-dimensional problem including Bundle detection~\cite{li2017neural}, bundle completion~\cite{bai2019personalized}, Bundle Naming~\cite{bai2019personalized} and Bundle Recommendation~\cite{pathak2017generating}. Bundle completion aims at creating bundles dynamically based on the current session. However, this work focuses on creating static bundles (can be updated with time) but does not recommend session-based bundles to consumers. Extant literature has multiple articles on Bundle recommendation, a step after identifying good bundles to recommend to consumers. In this work, We focus on creating new bundles. It is similar to bundle detection bundle as it detects potentially popular bundles but does not rely only on co-purchases and co-views. We introduce a new dimension to product bundling, that is, identifying the popularity of a bundle. 

\subsection{Popularity Metrics in Recommender Systems}\label{sec:recommenderSystems}
Any recommender system aims at recommending relevant items to its consumers. The relevance of the items depends on the content of the item and consumers preferences towards the item. To integrate both these factors, existing recommender systems use past consumer behavior, e.e. purchases, views, likes, shares, comments, to identify if an item could be appealing to other consumers as well. But how to find define if an item is ``appealing``? Number of views/clicks/past purchases are the most commonly used metrics~\cite{song2020towards}. Nayak \textit{et al.}~\cite{nayak2023news} discusses advantages and disadvantages of different commonly used metrics. 

Market Basket problem is similar to bundle detection problem. It is aimed at identifying items that are purchased together. Sequential basket problem aims at creating dynamic basket based on current consumer session~\cite{gholami2022parsrec}. Along with past click behavior, they also consider diversity of items. Coverage is well-studied metric for recommender systems~\cite{puthiya2016coverage}. Coverage is defined as how many types of items are covered in the list of recommended items. Diversity as a metric is defined as the pairwise dissimilarity between items in the list of recommendations~\cite{castells2021novelty}. We adopt coverage and diversity for using in the context of bundle popularity.  Note that metrics to identify popular bundles are different from evaluation metrics used for measuring the performance of a given recommender systems e.g. hit rate, recall, accuracy. 

We add to the literature on recommender systems by proposing new metrics for predicting the quality of a bundle. We further use the proposed metrics to build product bundles. We also add to the literature on bundle recommendations as new recommendation engines can be built around the proposed metrics instead of currently used number of clicks information.

\subsection{Bundle Generation}\label{sec:litBundleGeneration}
Extant literature on product bundling in scarce, hence the field is in nascent stage. This is most likely due to the lack of open-source corroborated as in~\cite{sun2022revisiting}. As discussed in~\cite{avny2022bruce}, bundle generation is an exciting topic to explore.

We extend the extant literature on product bundling by providing a holistic methodology for bundle generation. To the best knowledge of the authors, this is one of the first works in exploring different metrics for defining the popularity of a bundle and building a product bundling methodology based on it. We are also one of the first works in using item embeddings for sampling items to create new bundles.

\subsection{Embeddings}\label{sec:embeddings}
We use embeddings to condense information about a game in a n-Dimensional dense vector. We use two sets of embeddings - sentence embeddings and context embeddings. Sentence embeddings is a well studied in the field of Natural Language Processing (NLP). Sentence embeddings use word embeddings, revolutionized after transformers~\cite{vaswani2017attention}. Examples of sentence embeddings include Universal Sentence Embeddings (based on transfer learning and \cite{cer2018universal}), SBERT-WK (dissecting BERT layers, \cite{wang2020sbert}). In this work, we use SBERT~\cite{reimers2019sentence} and  Fasttext~\cite{mikolov2018advances} for sentence embeddings.

Brosh~\textit{et al.} \cite{avny2022bruce} create embeddings for consumers and items using attention-based transformer models for bundle recommendation task. They use one hot vector for users and bundles and recommend bundles to users where user's context embeddings is closest to a bundle. They further use games information to add content to the bundle information. Motivated by this work, we  use embeddings to represent games and bundles. However, since our work focuses on bundles generation, we do not use users' embeddings.

We contribute to the literature of sentence embeddings by exploring a new application of using embeddings. We use embeddings to identify the popularity of a bundle. We also use embeddings in sampling games to change existing bundles. Our work is one of the first in exploring the use of embeddings for product bundling and provides directions for future research. Next, we discuss the bundling problem before introducing the proposed methodology and its results.

\section{Problem Formulation}\label{sec:problemFormulation}
In this Section, we provide a detailed discussion on formulating product bundling. Before discussing the problem in detail, we discuss the Steam Data used in this paper.

\subsection{Data}\label{data}
\begin{figure}
	\centering
	\begin{subfigure}[b]{0.23\textwidth}
		\centering
		\includegraphics[width=\textwidth]{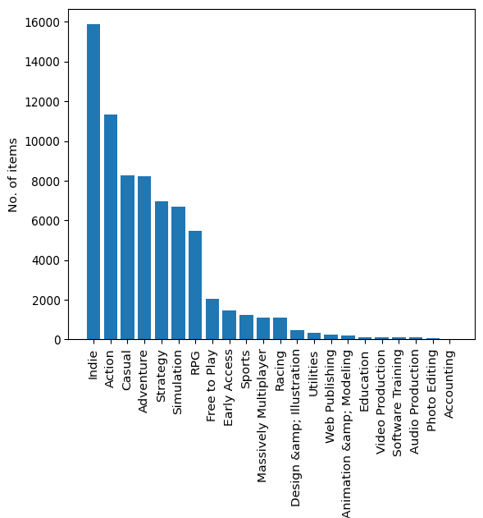}
		\caption{Distribution of Genres}
		\label{fig:genres_dist}
	\end{subfigure}
	\begin{subfigure}[b]{0.23\textwidth}
		\centering
		\includegraphics[width=\textwidth]{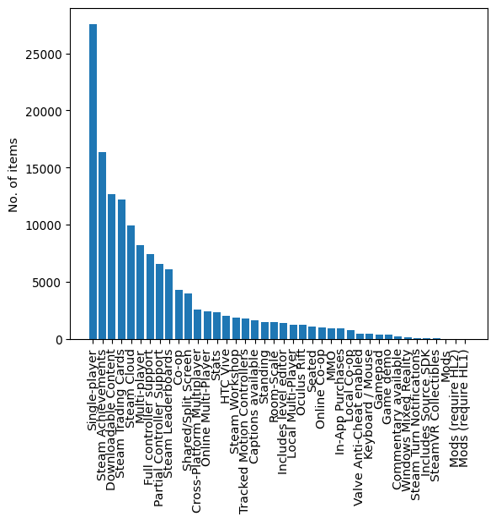}
		\caption{Distribution of Tags}
		\label{fig:tags_dist}
	\end{subfigure}
	\begin{subfigure}[b]{0.23\textwidth}
		\centering
		\includegraphics[width=\textwidth]{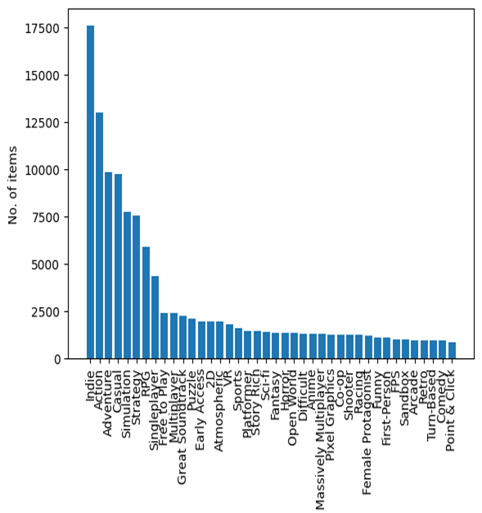}
		\caption{Distribution of Specs}
		\label{fig:specs_dist}
	\end{subfigure}
	\caption{Game Features in $Large$ Dataset}
	\label{fig:tagsGenresSpecs}
\end{figure}

We use open-source dataset from Steam Video Games to build and test our methodology on product bundling. We combine two datasets from Steam. We call these datasets as $Large$ and $Small$ dataset. Most of the analysis in extant literature is based on $Small$ dataset. $Large$ dataset contains information on 35000 games and purchase history of 83000 users. Game information includes the title (name), genres, tags, specifications provided by the game author. tags, genres and specifications are used by users to understand the type of the game. Distribution of top ten tags, genres and specifications is shown in Figure~\ref{fig:tagsGenresSpecs}. It also includes game price, launch time and developer company of the game. The purchase history of the users in pre-training dataset contains approximately 10000 games. This indicates that approximately 70\% of the games were never played. In our study, we do not consider this unplayed games and focus our analysis on the 10000 games that were played at-least by the users in $Large$ dataset. Pre-train dataset also contains information on the lifetime playtime for a game by a user. 

\begin{figure}[!h]
	\centerline{\includegraphics[width=0.8\linewidth]{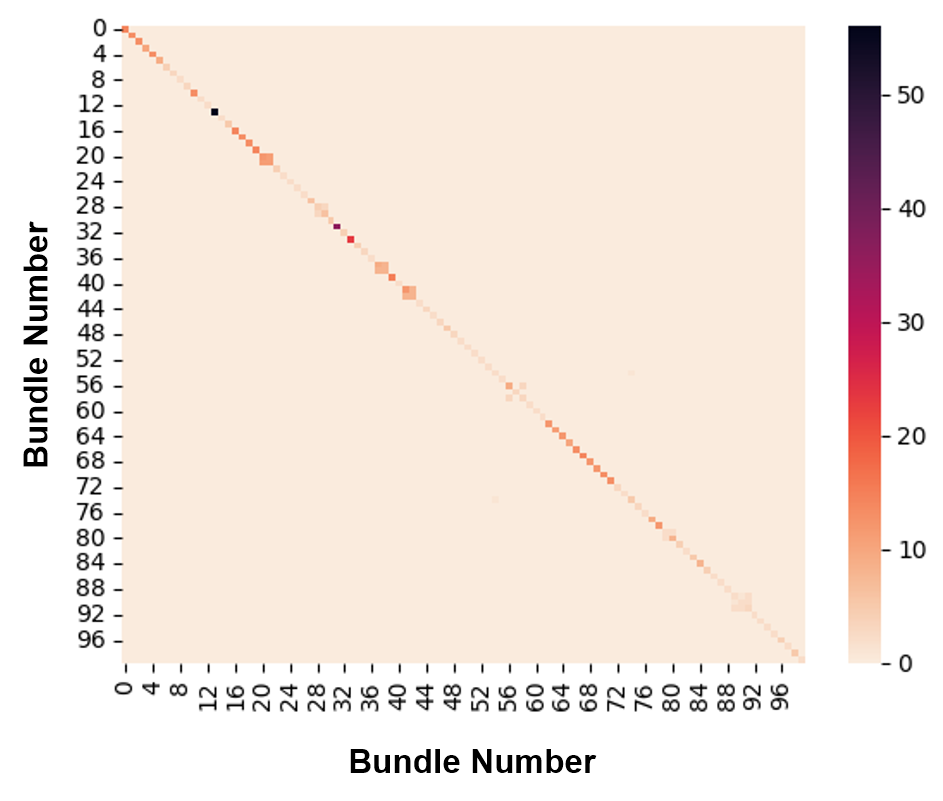}}
	\caption{Intersection of Games in Bundles}
	\label{fig:intersection}
\end{figure}

However, the $Large$ dataset does not contain information about the bundles. For bundles, we consider the $Small$ dataset. It includes information on 23000 users, 615 bundles and 2819 games. The dataset does not provide the rationale behind how these 2819 games are selected for bundling. Extant research on bundle recommendation use the existing bundles for evaluating their model performances. Figure~\ref{fig:intersection} shows the intersection of games in any two bundles (number of common games in bundles). It underlines an important aspect as how the bundles are made, that is, it shows most bundles are constructed such that each bundle has a different set of games and very few intersection of games exists among bundles. For clarity, we include randomly selected 100 bundles in Figure~\ref{fig:intersection}.

Users in the $Small$ dataset are sampled randomly from the users in the $Large$ dataset. Note that the $Small$ dataset only contains user-bundle interaction while the $Large$ dataset contains user-item interaction only. User-item interaction can be imputed from the user-bundle interaction as each bundle contains a set of items included in that bundle. Thus, expanding user-bundle interactions can provide user-item interaction. $Small$ dataset also does not contain playtime information about the user-game interaction. The summary statistics for the $Large$ and $Small$ dataset is shown in Table~\ref{table:data_summary}. For Brevity, we have shown playtime in minutes as $log (playtime)$, encoded sentiments as Very Positive = 5, Positive = 3, Negative = 2 and Very Negative = 1. Age of a game is calculated from their published date to August 30, 2023.
\begin{table}[!h]
	\centering
	\caption{Data Summary for $Large$ and $Small$ dataset}
	\label{tab:data_summary}
	\begin{tabular}{l|c}
		\hline
		\multicolumn{2}{c}{(\textbf{a}) $Large$ Dataset}\\
		\hline
		Features & (mean, median, max, std)\\
		\hline
		$log$ (Playtime of Games in Bundle) & (6.95,7.89,20.48,4.21) \\
		Number of downloads of Games & (420.4,33,43776,1681) \\
		Age of a Game (in years)& (8.92, 7.66, 40.6, 4.54) \\
		Price of a Game & (9.89,6.99,771,12.8) \\
		Sentiment & (3.68,4,5,1.61) \\
		\hline
		\multicolumn{2}{c}{(\textbf{b}) Bundles in $Small$ Dataset}\\
		\hline
		Features & (mean, median, max, std)\\
		\hline
		Number of Games in a Bundle & (5.73, 3, 89, 8.1) \\
		Number of Purchases & (0.0 , 188.9 , 14000.0 , 1175.05) \\
		Price of a Bundle & (47.01,24.98,999.98, 78.94) \\
		Discount of a Bundle (in \%)& (24.46,20,92,16.92) \\
		$log$ (Playtime of Games in Bundle) & (4.79,3.21,20.48,5.09) \\
		Number of downloads of Games & (438.1,2,43776,2040) \\
		\hline
		
	\end{tabular}
\end{table}

We use Steam dataset because it meets the requirement information in the data -- consumers past purchase information, content information about the items and existing bundles curated by the firm which allows to construct a model and provide a baseline for testing the performance of our proposed methodology for bundle creation. Next, we discuss as how to define the quality of a bundle from the perspective of the likelihood of  a bundle becoming popular.

\subsection{Bundle Popularity Metrics}\label{sec:modelBundlePopularity}
Before creating a new bundle, we need to understand if a new bundle has the potential to become a popular bundle. Extant literature considers clicks among recommended bundle as a metric to identify popular bundles. However, as discussed in Section~\ref{sec:introduction}, this metric suffers from multiple disadvantages and need further research to identify popular bundles. In this Section, we introduce different metrics with their rationale and limitations. 

Let $x_{b,i}=1$ if an game $i$ is present in bundle $b$ where $B=\{b_1,b_2,...\}$ is set of all the Bundles. Let $\hat{U}$ be the set of users in pre-train dataset and $U$ be the set of users in the training dataset. Total playtime of a game is given by $P_i = \sum_u^{\hat{U}}p_{u,i}$ where $p_{u,i}$ is the total time (in minutes) spent by user $u$ on game $i$. Let $B_b={i_1,i_2,...}$ be the set of games in Bundle $b$ and $G_u={i_1,i_2,..}$ be the set of games purchased by user $u$.

The metrics proposed in this paper to define the quality (or popularity) of a game include:
\subsubsection{Explicit Number of Purchases ($P_{eb}$)}
\begin{figure}
	\centering
	\begin{subfigure}[b]{0.23\textwidth}
		\centering
		\includegraphics[width=\textwidth]{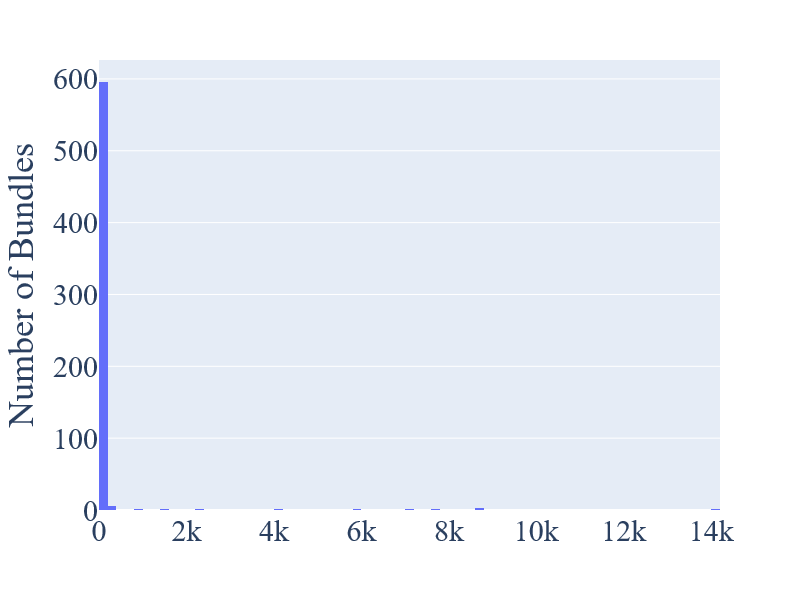}
		\caption{Histogram}
		\label{fig:m1Histogram}
	\end{subfigure}
	\begin{subfigure}[b]{0.23\textwidth}
		\centering
		\includegraphics[width=\textwidth]{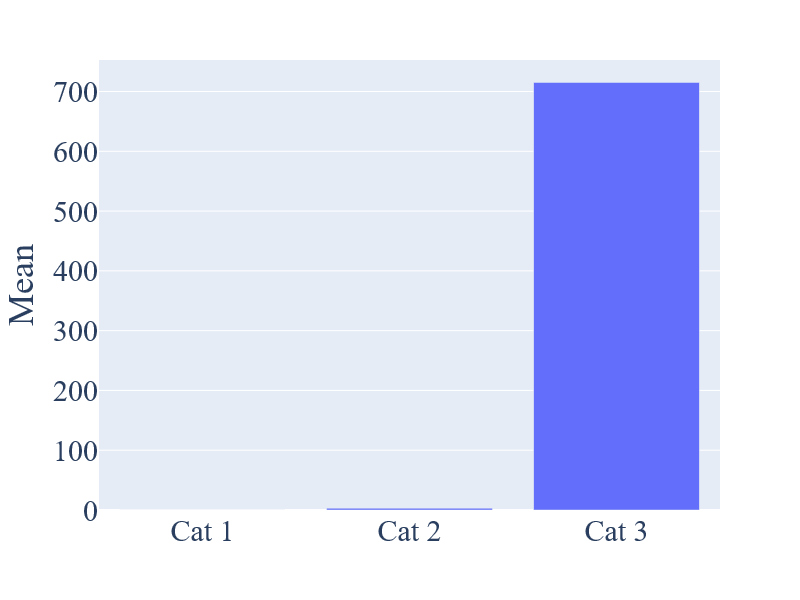}
		\caption{Categorical Distribution}
		\label{fig:m1Dist}
	\end{subfigure}
	\caption{Distribution of Purchases in $Small$ Dataset}
	\label{fig:bundlesmallpurchase}
\end{figure}

Explicit number of purchases is the count of the users in $Small$ dataset that purchase a bundle $b$. We call it explicit because this is obtained directly from the dataset provide (only $Small$ dataset has bundle information). It is shown by Equation~\ref{eq:epurchase} where $y_{u,b}=1$ if user $u$ purchases a bundle $b$. The distribution for the number of times a bundle is purchased in $Small$ dataset is shown in Figure~\ref{fig:bundlesmallpurchase}. This metric is synonymous to the click popularity metric used in extant literature. However, it could be an artefact of the recommendation engine that recommends the bundles as discussed in Section~\ref{sec:introduction}.
\begin{equation}\label{eq:epurchase}
	P^e_{b}=\sum_u^{U}y_{u,b} \quad \forall b\in B
\end{equation}

\subsubsection{Implicit Number of Purchases ($P_{mb}$)}
\begin{figure}
	\centering
	\begin{subfigure}[b]{0.23\textwidth}
		\centering
		\includegraphics[width=\textwidth]{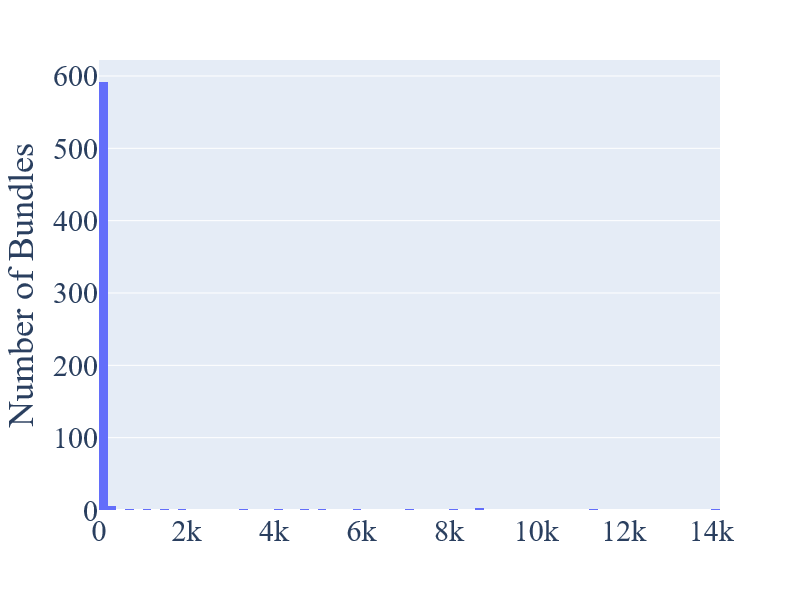}
		\caption{Histogram}
		\label{fig:m2Histogram}
	\end{subfigure}
	\begin{subfigure}[b]{0.23\textwidth}
		\centering
		\includegraphics[width=\textwidth]{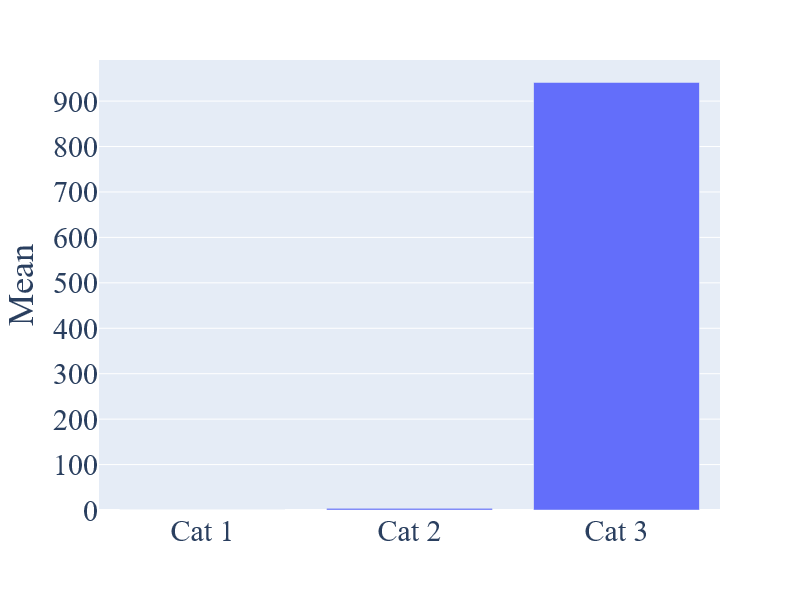}
		\caption{Categorical Distribution}
		\label{fig:m2Dist}
	\end{subfigure}
	\caption{Distribution of Purchases in $Large$ dataset}
	\label{fig:bundlelargepurchase}
\end{figure}

Implicit number of purchases is the count of the users in $Large$ dataset that purchase at least 80\% of the games in bundle $b$. We call it implicit because we do not have explicit information about the purchase of a bundle or intent of the bundle. Therefore, we approximate it by the purchase history of individual games. We use this metric to estimate how popular a set of games could be. It can be useful for businesses where there is no bundle data yet and it can be used to create new bundles. It is shown by Equation~\ref{eq:ipurchase}. The distribution for the number of times a bundle is purchased in $Large$ dataset is shown in Figure~\ref{fig:bundlelargepurchase}. W. This metric is synonymous to the existing methods of bundle generation that bundle items based on past co-purchases.
\begin{equation}\label{eq:ipurchase}
	P_{mb}=\sum_u^{\hat{U}}y_{u,b} \,\, \text{where} \,\, y_{u,b}=1 if \vert G_u \cap B_b \vert > 0.8 \vert B_b \vert
\end{equation}

\subsubsection{Number of 0 Playtime Games, $N^0_b$}
Consumers are price sensitive, especially in buying product bundles. This is because discounts is one of the major attributes of bundles that incentives the consumers. A consumer may not buy a bundle if the bundle includes items which are not interesting to them. So a consumer may avoid a bundle even at discounted price to avoid paying for an item they do not want. To capture this human behavior, we propose $N^0_b$ in Equation~\ref{eq:playtime0} where $I(x)$ is an indicator function. This metric may not be useful for new games as newer games will have 0 playtime.
\begin{equation}\label{eq:playtime0}
	N^0_b=\sum_i^{B_b}I(P_i==0) \quad \forall b \in B
\end{equation}

\begin{figure}
	\centering
	\begin{subfigure}[b]{0.23\textwidth}
		\centering
		\includegraphics[width=\textwidth]{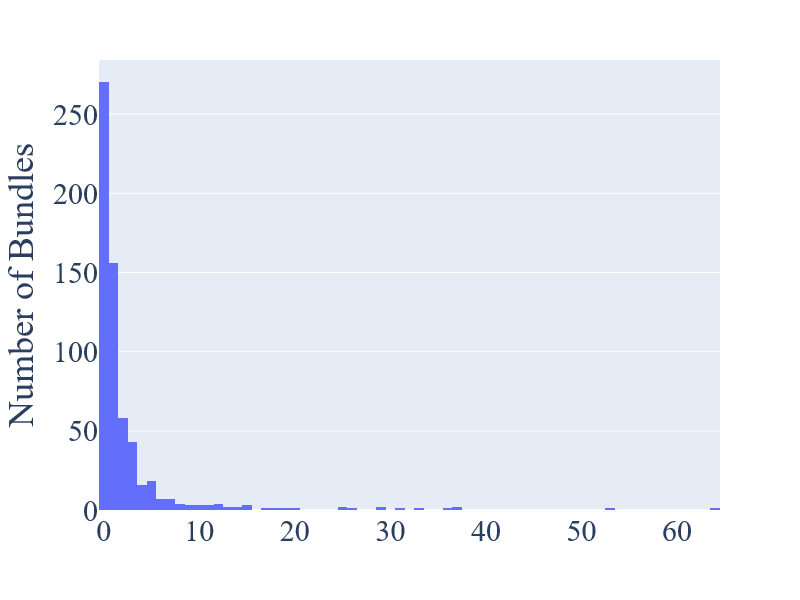}
		\caption{Histogram}
		\label{fig:m3Histogram}
	\end{subfigure}
	\begin{subfigure}[b]{0.23\textwidth}
		\centering
		\includegraphics[width=\textwidth]{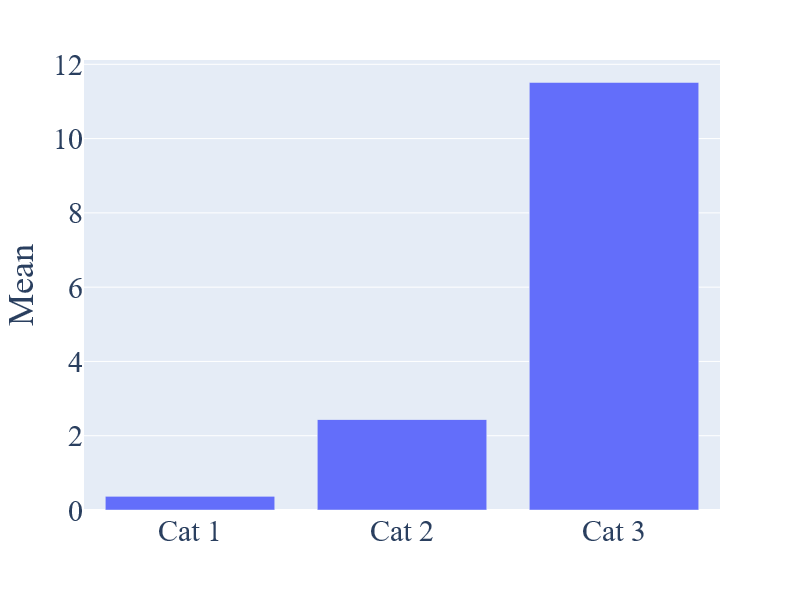}
		\caption{Categorical Distribution}
		\label{fig:m3Dist}
	\end{subfigure}
	\caption{Distribution of Number of Games with 0 Playtime}
	\label{fig:zeroPLaytime}
\end{figure}

\subsubsection{Total Playtime of Bundle, $P^B_b$}
We postulate that including good items in a bundle may create bundles interesting to consumers as other consumers have previously liked the items in this bundle. Total playtime of a bundle is given by $P^B_b$ in Equation~\ref{eq:totalplaytime}. Similar to $N^0_b$, this metric is not useful for new products. This may also lead to just bundle popular items in a bundle, which in turn could lead to revenue loss for the firm (as popular products would have sold even without the discounts provided in bundles).
\begin{equation}\label{eq:totalplaytime}
	P^B_b=\sum_i^{B_b}P_i \quad \forall b \in B
\end{equation}

\begin{figure}
	\centering
	\begin{subfigure}[b]{0.23\textwidth}
		\centering
		\includegraphics[width=\textwidth]{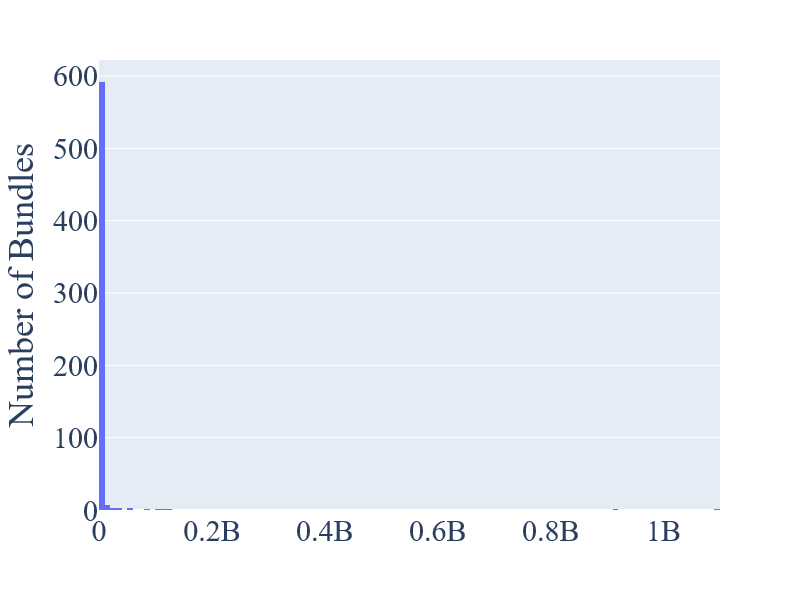}
		\caption{Histogram}
		\label{fig:m4Histogram}
	\end{subfigure}
	\begin{subfigure}[b]{0.23\textwidth}
		\centering
		\includegraphics[width=\textwidth]{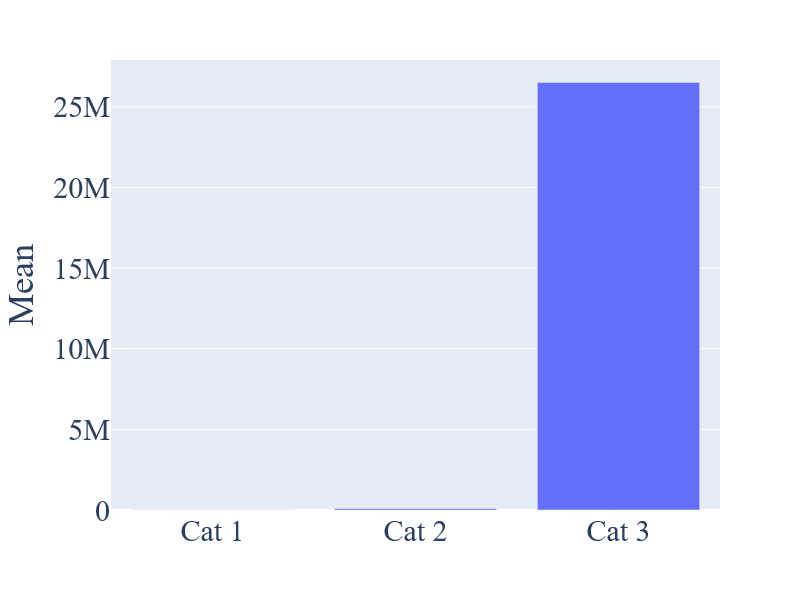}
		\caption{Categorical Distribution}
		\label{fig:m4Dist}
	\end{subfigure}
	\caption{Distribution of Playtime (in minutes) of Games in Bundles}
	\label{fig:totalPlaytime}
\end{figure}

\subsubsection{Diversity in a Bundle, $D_b$}
\begin{figure}
	\centering
	\begin{subfigure}[b]{0.23\textwidth}
		\centering
		\includegraphics[width=\textwidth]{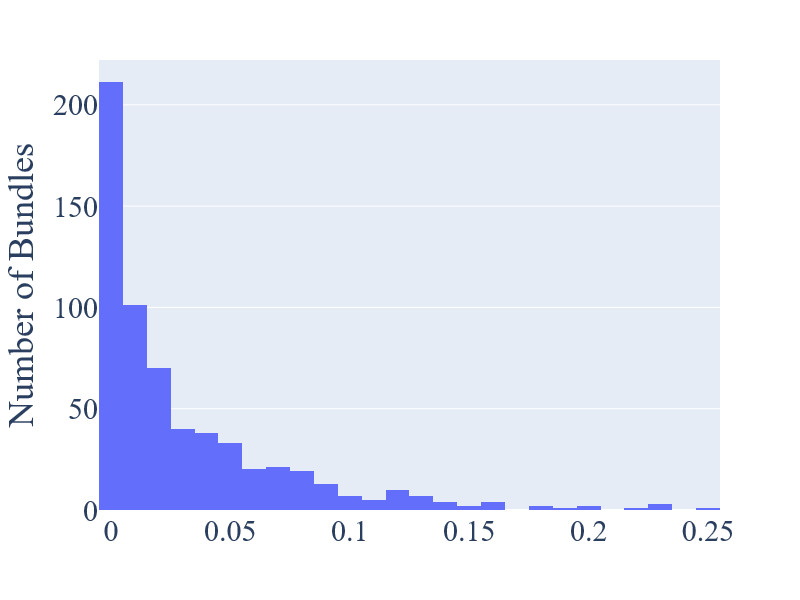}
		\caption{Histogram}
		\label{fig:m5Histogram}
	\end{subfigure}
	\begin{subfigure}[b]{0.23\textwidth}
		\centering
		\includegraphics[width=\textwidth]{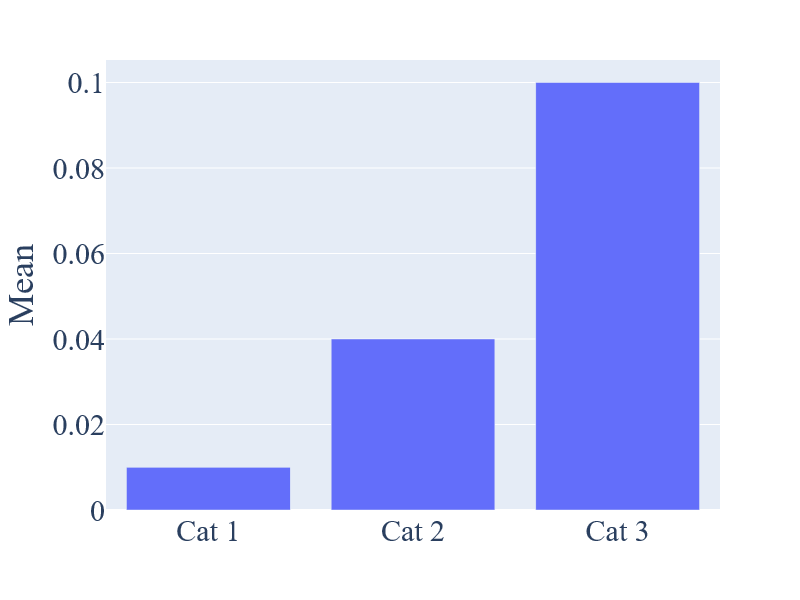}
		\caption{Categorical Distribution}
		\label{fig:m5Dist}
	\end{subfigure}
	\caption{Distribution of Bundle Diversity}
	\label{fig:diversity}
\end{figure}

Bundles can be made with different strategy. One strategy could be to include similar product (e.g. games or books) and other strategy could be to include variety of items in a bundle (e.g. news). To understand the relationship between games in a bundle, we use average cosine similarity a we define diversity as $D_b$ in Equation~\ref{eq:diversity} where $cos(i,j)$ is the cosine similarity between items $i$ and $j$. Distribution of diversity in bundles is shown in Figure~\ref{fig:diversity}. However, high diversity may or may not lead to popularity of bundles, thus it is important to for firms to consider which diversity works for business.
\begin{equation}\label{eq:diversity}
	D_b = 1-\frac{\sum_{i,j}cos(i,j\in B_b)}{\vert B_b^2\vert} \quad \forall b \in B
\end{equation}

\subsubsection{Coverage of Bundles, $C^B$}
Coverage is defined as how many types of items are included in the list of recommendations. In this paper, we adopt this concept to provide overall coverage of the set og bundles $B$. Equation~\ref{eq:coverage} defines coverage $C^B$. The value for coverage for the existing bundles is 0.3. Higher coverage relates to creating bundles not just from the popular items but creating diverse bundles so that it covers wider range of items. This will help the firm in exposing items to consumers which are not ``popular'' which may be due to the artifact of the recommendation engine. 
\begin{equation}\label{eq:coverage}
	C^B = 1-\frac{\sum_{b\in B}\sum_{i,j}cos(i,j\in B_b)}{\sum_{b \in B}\vert B_b^2\vert}
\end{equation}

Statistics of the proposed metrics for the current bundles (615 in $Small$ dataset) is shown in Table~\ref{table:metricStatistics}. For diversity and Coverage, we use metrics obtained across different embeddings. We model product bundling as classification problem, thus we categorize bundles as Unpopular ($Cat 1$), Popular ($Cat 2$) and Very popular ($Cat 3$) bundles based on percentiles. On the left part of Table~\ref{table:metricStatistics}, we provide the count of bundles in each of the three categories.

\begin{table}[!h]
	\centering
	\caption{Metrics Statistics for Current Bundles}
	\label{table:metricStatistics}
	\begin{tabular}{c|ccc|ccc}
		\hline
		Metrics& \multicolumn{3}{c}{Statistics} & \multicolumn{3}{c}{Number of Bundles}\\
		& \multicolumn{3}{c}{(mean, median, max, std)} &\multicolumn{3}{c}{($Cat 1$, $Cat 2$, $Cat 3$)} \\
		\hline
		$P_{eb}$&\multicolumn{3}{c}{(0.0 , 142.38 , 14000.0 , 1011.8)}&\multicolumn{3}{c}{(369,123,123)}\\
		$P_{mb}$&\multicolumn{3}{c}{(0.0 , 188.9 , 14000.0 , 1175.05)}&\multicolumn{3}{c}{(378,114,123)}\\
		$N_{b}^0$&\multicolumn{3}{c}{(1.0 , 2.3 , 64.0 , 5.69)}&\multicolumn{3}{c}{(426,101,88)}\\
		$log(P_{b}^B)$&\multicolumn{3}{c}{(9.54,15.49, 20.81,17.88)}&\multicolumn{3}{c}{(369,123,123)}\\
		$D_b$&\multicolumn{3}{c}{(0.01 , 0.05 , 1.0 , 0.11)}&\multicolumn{3}{c}{(307,154,154)}\\
		$C^B$&\multicolumn{3}{c}{(0.21 , 0.21 , 0.62 , 0.13)}&\multicolumn{3}{c}{-}\\
		\hline
		
	\end{tabular}
\end{table}

We can use the proposed metrics to estimate the popularity of a bundle, existing or new. Next, we discuss our proposed approach for new bundles creation.

\subsection{Product Bundling}\label{sec:modelProductBundling}
Product bundling aims at creating high quality bundles that serve business purpose, e.g. increasing purchases, increasing revenue or consumer engagement. We use the metrics discussed in Section~\ref{sec:modelBundlePopularity} as surrogate to increasing revenue or consumer engagement. A bundle $b$ is a sub-set of items $i$ in $I$. These items are grouped to form a bundle using different strategies. Conceptually, a utility function shown in Equation~\ref{eq:utility} maps how well a bundle or combined characteristics of items in the bundle interests consumers. In this paper, we use machine learning models in Section~\ref{sec:predictBundlePopularity} to estimate the utility of a bundle based on the metrics introduced in this paper. 
\begin{equation}\label{eq:utility}
	U_b = g(i_1,i_2,... \in B_b)\quad \forall b \in B
\end{equation}

Methodology proposed in this paper can be extended to create new bundles using co-purchase, co-views or past consumer purchases. In this approach, co-purchases or co-views can be loosely considered as existing bundles. We can build the machine learning model to to understand how the relationship between items leads to popular bundles and use the model to create new bundles. However, we leave it as a future direction and focus our analysis on Steam dataset which provides information about ground truth bundles.

\section{Proposed Methodology}\label{sec:proposedMethodology}
In this section, we provide a layout for creating new bundles. First, we create embeddings for different games. Then we construct machine learning models to estimate bundle quality (likelihood of being popular). Finally, we use these model to optimize bundles through various strategies. We explain the steps next.

\subsection{Game Embeddings}\label{sec:gameEmbeddings}
\begin{figure}[!h]
	\centerline{\includegraphics[width=\linewidth]{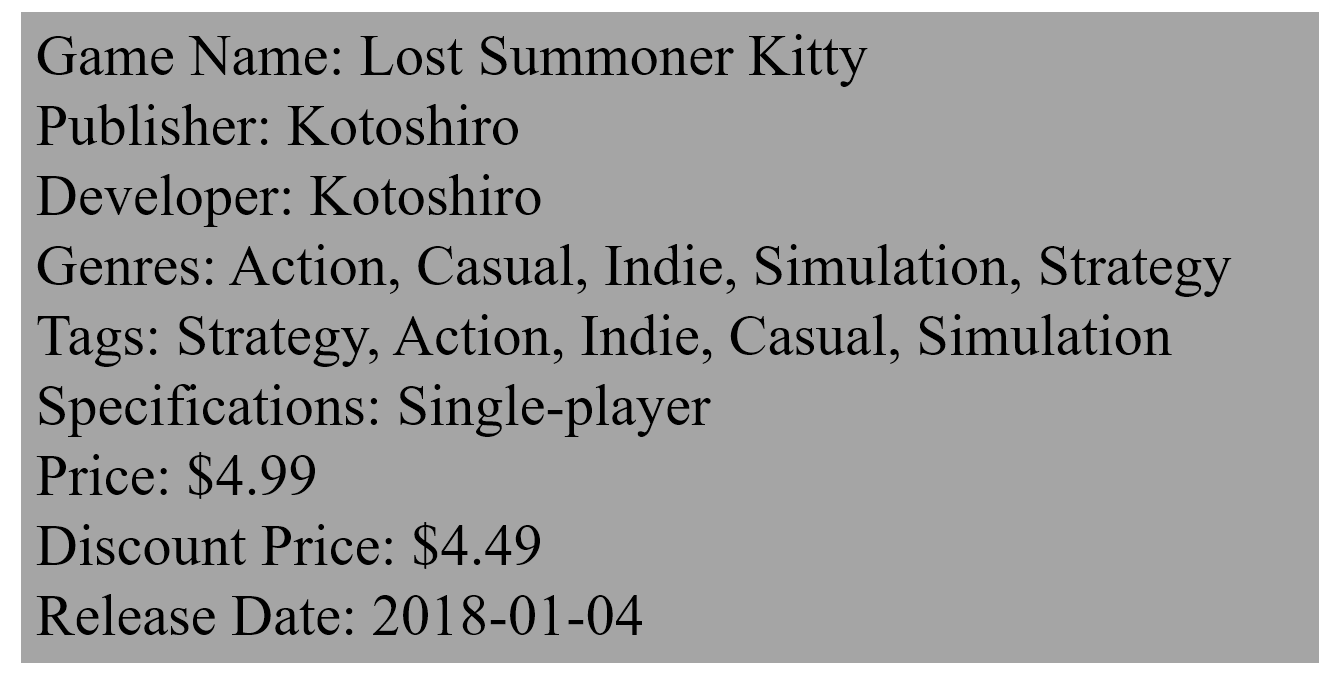}}
	\caption{Example of a Game and its Available Information}
	\label{fig:gameExample}
\end{figure}

We use two types of embeddings for creating embeddings of games -- sentence embeddings and context embeddings. In sentence embeddings, we use the information provided by the game developers and assume that the information provided by them is accurate representation of the game. The developers provide title of the game, genres of the game they believe it belong to, tag words for searching the game and game specifications (specs). An example of the a game with all its information is shown in Figure~\ref{fig:gameExample}. We use title only, title+tags, title+genres, title+specs and title+tags+genres+specs to generate sentence embeddings. We refer readers to Seo et al~\cite{seo2022ta} for further reading on sentence embeddings. Conceptually, sentence embeddings use word embeddings to condense information from all the words in a sentence. Word embedding models use contextual information in a sentence to predict the next word. Both word embeddings and sentence embeddings are studied extensively in literature. Specifically, we use SBERT~\cite{reimers2019sentence} and Fasttext sentence embeddings. SBERT uses attention-based transformer model and if computationally expensive and slow. SBERT provides a 384-Dimensional vector for sentence embeddings. Fasttext extends Word2Vec model~\cite{church2017word2vec} and is computationally fast. It provides a 300-Dimensional vector for sentence embeddings.

For context embeddings, we use Prod2Vec~\cite{turgut2023prod2vec} and MetaProd2Vec~\cite{vasile2016meta}. Both the models are extension of Word2Vec model. We refer readers to Word2Vec~\cite{church2017word2vec} for further reading on Word2Vec. Conceptually, Prod2Vec, $P2V$ models products purchased by a consumer as a sentence and each item as a word. It then uses Word2Vec model to find embeddings for the word (item in Prod2Vec). MetaProd2Vec, $M-P2V$ extends Prod2Vec as includes item information while creating the embeddings. Next we discuss how we use these embeddings.

\subsection{Predicting Bundle Popularity}\label{sec:predictBundlePopularity}
For bundle generation, first we need to estimate how popular/non-popular a bundle can be, based on the games included in that bundle. We use a machine learning model shown in Equation~\ref{eq:ML} to estimate if the bundle could be popular or non-popular ($f(X)$ is the functional form of data $X$). Due to skewness in the dataset, we model the problem of identifying the popularity of a bundle as classification problem. The distribution of number of bundles in each category is shown in respective Sections of popularity metrics. $X$ for building the model includes game embeddings $embeddings_i$, price of bundle $price_b$, age of bundle $age_b$ and discount offered $discount_b$. We use mean of $embeddings$ of games in a bundles. $X$ also includes vectors for categorical information of games in the bundle, that is, top tags of the games $tags_b$, top genres of the games $genres_b$, top specs of the games $specs_b$ and sentiment of the game $sentiment_b$. We also include aggregated playtime information of the games in the bundle, e.g., total purchases of the games $totalPurchase_b$ and total playtime per downloads of the games $playTimePerDownload_b$. Note that we carefully select $X$ for building models for different metrics to avoid adding metrics in $X$ itself e.g. $totalPurchase_b$.
\begin{equation}\label{eq:ML}
	\frac{P_b}{1-P_b} = f(X) \quad \forall b \in B
\end{equation}

One of the major limitation of this work is that the number of bundles ($\vert B \vert$ = 615) is small. Thus, constructing complex non-linear machine learning models are prone to overfitting. To counter this, we use extensive experimentations and use simpler models for Equation~\ref{eq:ML}. All the results discussed in this paper hereon are built using logistic regression, thus, $f(X)$ is a linear function. Moreover, to reduce the size of the model, we use 2-Dimensional embeddings of games in our logistic regression model. We also provide the github repository for replicating the results where readers can use different machine learning models along with logistic regression. All the models in this study were built on Windows 10 platform, i7 Intel processor with 16 GB RAM. We use Python 3.8 and open-source python libraries for all the analysis.

\subsection{New Bundle Creation}\label{sec:bundleGeneration}
\begin{figure}[!h]
	\centerline{\includegraphics[width=\linewidth]{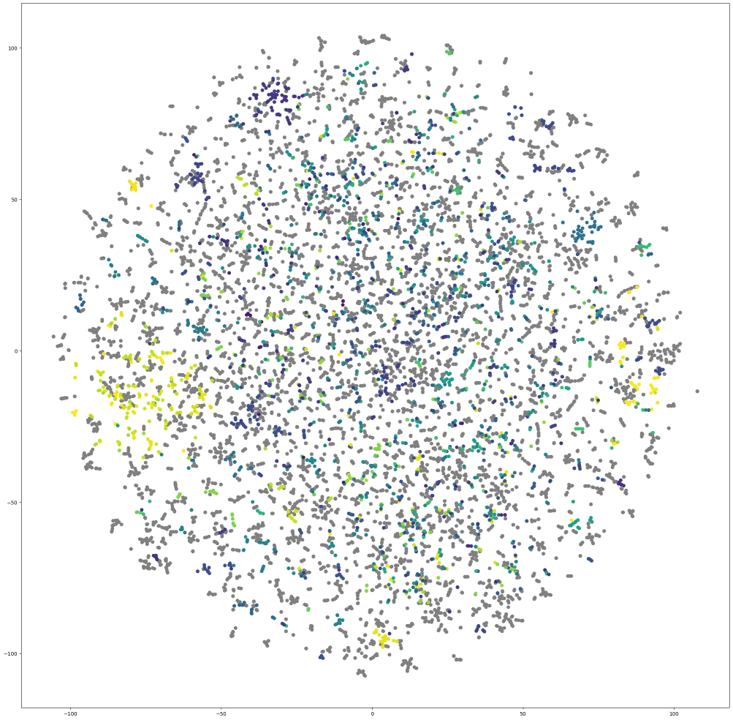}}
	\caption{Representing Games in 2-D Latent Space}
	\label{fig:allGames}
\end{figure}

Bundle generation aims at finding items, $i\in I$ to make a potentially popular bundle. In this work, we use distance-based metrics to sample items from $I$ to create new bundles. We use two types of embeddings to calculate distance for sampling. First, we use different combinations explicit textual features of the items (e.g. title, genres, tags and specifications) to calculate sentence embeddings. Second, we use purchase behavior from $Large$ dataset to find embeddings of items using Prod2Vec and MetaProd2Vec. An example of the sentence embeddings using SBERT when sentence is give by title+genres is shown in Figure~\ref{fig:allGames}. Figure~\ref{fig:allGames} shows 10000 games  from $Large$ dataset, dimensionally reduced to two dimensions using UMAP dimensionality reduction model. Figure~\ref{fig:allGames} also shows the games that are included in different bundles. The games are colored by the bundle numbers which shows that similar games are included in a bundle. 

To create new bundles, we use four strategies to create new bundles (1) inserting a game in existing bundle (2) exchanging games from existing bundles (3) deleting games from existing bundles (4) creating bundles from scratch using seed games.  First, we sample a game based on embeddings and perform these operations to create new bundles. To check if the new bundle has the potential to be popular or not, we test the performance of the new bundle on the proposed metrics using machine learning model (in Section~\ref{sec:predictBundlePopularity}). The pseudo-code for bundle creation is shown in Table~\ref{table:pseudocode}.
\begin{table}[!h]
	\centering
	\caption{Pseudocode for Creating New Bundles}
	\label{table:pseudocode}
	\begin{tabular}{ll}
		\hline
		\multicolumn{2}{c}{Pseudocode}\\
		\hline
		Steps & \textbf{Details}\\
		1 & Pick the appropriate embeddings from Section~\ref{sec:gameEmbeddings}\\
		2 & Represent each game as embeddings from Step 1.\\
		3 & Select a Bundle at random\\
		4 & Remove a game with a probability $p$\\
		5 & \begin{tabular}{@{}l@{}}Select a game probabilitically based on distance from\\ the bundle embeddings centroid\end{tabular}\\
		6 & \begin{tabular}{@{}l@{}}Check if adding the game in Step 5 in the bundles\\ improves bundle metrics\end{tabular} \\
		7 & Repeat Steps 3 - 6 until desired quality bundles in obtained\\
		\hline
		
	\end{tabular}
\end{table}

Next, we discuss the results for the proposed methodology and compare it with the performance of the existing bundles.

\section{Results and Discussion} \label{sec:results}
We summarize the performance of the different models embeddings
for the popularity metrics in Table~\ref{table:model_results}. We use “all" for embeddings
with title + tags + genres + specs.We present the results for Random
Forest model.
\begin{table}[!h]
	\caption{Model Performance for Popularity Metrics}
	\label{table:model_results}
	\begin{tabular}{l|ccccc}
		\hline
		& \multicolumn{5}{c}{Metrics (AUC, F1-score) x 100}\\
		\hline
		& $P_{eb}$& $P_{mb}$ & $N_{b}^0$ & $P_{b}^B$ & $D_b$\\
		\hline
		SBERT+&&&&&\\
		title&(78,64)&(81,72)&(92,79)&(81,72)&(83,69)\\
		+tags&(78,80)&(79,78)&(91,82)&(79,77)&(82,70)\\
		+genres&(80,64)&(79,72)&(91,81)&(79,72)&(76,57)\\
		+specs&(79,55)&(80,85)&(91,86)&(80,73)&(75,58)\\
		all&(78,54)&(81,85)&(92,80)&(81,73)&(74,59)\\
		\hline
		Fasttext&&&&&\\
		title+&(79,57)&(90,85)&(99,76)&(83,69)&(78,62)\\
		+tags&(79,58)&(79,85)&(90,80)&(82,69)&(84,64)\\
		+genres&(79,53)&(78,86)&(92,82)&(76,57)&(77,63)\\
		+specs&(78,53)&(81,85)&(91,74)&(81,72)&(75,58)\\
		all&(78,54)&(80,85)&(92,80)&(80,74)&(74,59)\\
		\hline
		P2V&(81,56)&(79,86)&(91,87)&(69,22)&(79,68)\\
		\hline
		M-P2V&&&&&\\
		+title&(78,51)&(80,85)&(91,87)&(69,22)&(79,68)\\
		+title+tags&(79,59)&(81,85)&(90,91)&(68,23)&(75,76)\\
		+title+genres&(79,54)&(80,83)&(91,91)&(69,24)&(77,71)\\
		+title+specs&(79,53)&(80,85)&(91,91)&(69,24)&(75,70)\\
		+all&(80,51)&(74,86)&(92,91)&(68,28)&(75,70)\\
		\hline
		
	\end{tabular}
\end{table}

Results indicate that SBERT+title+genre consistently performs
well across all the metrics discussed in this paper.We use this embeddings
for bundle creation and show the results in Table 4. Content
aware embeddings perform better than context aware embeddings
on average, therefore $M-P2V$ outperforms $P2V$. Experiments show 
that the quality of individual games (average playtime and average
number of downlaods) are also important for determining the
quality of a bundle. Thus it is important to include popular games
in a bundle. Common tags and genres are also useful in predicting
the popularity of bundles, which highlights the importance of
providing accurate game description by game developers.
For creating new bundles, we follow three strategies discussed
in Section 3.3.We use the embeddings to sample the games. Results
for bundle creation is shown in Table 4. It shows the percentage
of the number of existing bundles we were able to upgrade from
Non-popular to Popular and Popular to Very Popular by using the
strategies. We show the 95\% bounds obtained from simulation runs.
Results indicate that carefully changing the games of existing bundles
can help in creating better bundles. Removing games did not
have much effect on the popularity of bundles. This is mainly because
number of games is not statistically significant in influencing
the popularity of a bundle.We present the result for different metric
in Table 4.

For creating new bundles, we follow four strategies to change the bundles by sampling games. We use the embedding space to sample the games. Results for the four strategies is shown in Table~\ref{table:new_bundles}. Table shows the percentage of the number of existing bundles we are able to move from Non-popular to Popular and Popular 2 to Very Popular. Results indicates that carefully changing games of existing bundles can help in creating better bundles. Removing games did not have much effect on the popularity of bundles. This is mainly because number of games is not statistically significant in influencing the popularity of a bundle. 
\begin{table}[!h]
	\caption{Comparative Performance of New Bundles}
	\label{table:new_bundles}
	\begin{tabular}{cl|ccc}
		\hline
		&Category Shift& \multicolumn{3}{c}{Strategies}\\
		\hline
	    && Replace& Add& Delete\\
		\hline
		$P_{eb}$&Unpopular$\rightarrow$Popular&(75\%-78\%)&(64\%-67\%)&(38\%-42\%)\\
		&Unpopular$\rightarrow$Very Popular&(67\%-84\%)&(57\%-62\%)&(13\%-17\%)\\
		&Popular$\rightarrow$Very Popular&(70\%-75\%)&(66\%-73\%)&(37\%-44\%)\\
		\hline
		$P_{mb}$&Unpopular$\rightarrow$Popular&(73\%-78\%)&(58\%-68\%)&(38\%-40\%)\\
		&Unpopular$\rightarrow$Very Popular&(40\%-45\%)&(19\%-23\%)&(14\%-18\%)\\
		&Popular$\rightarrow$Very Popular&(71\%-74\%)&(65\%-72\%)&(42\%-45\%)\\
		\hline
		$N_{b}^0$&Unpopular$\rightarrow$Popular&(72\%-78\%)&(58\%-67\%)&(38\%-42\%)\\
		&Unpopular$\rightarrow$Very Popular&(39\%-44\%)&(19\%-22\%)&(15\%-18\%)\\
		&Popular$\rightarrow$Very Popular&(68\%-75\%)&(67\%-72\%)&(42\%-45\%)\\
		\hline
		$P_{b}^B$&Unpopular$\rightarrow$Popular&(69\%-79\%)&(58\%-69\%)&(38\%-43\%)\\
		&Unpopular$\rightarrow$Very Popular&(41\%-45\%)&(21\%-24\%)&(15\%-19\%)\\
		&Popular$\rightarrow$Very Popular&(71\%-75\%)&(70\%-74\%)&(41\%-44\%)\\
		\hline
		 $D_b$&Unpopular$\rightarrow$Popular&(73\%-77\%)&(64\%-68\%)&(38\%-42\%)\\
		&Unpopular$\rightarrow$Very Popular&(41\%-46\%)&(20\%-24\%)&(13\%-19\%)\\
		&Popular$\rightarrow$ Very Popular&(68\%-75\%)&(67\%-73\%)&(39\%-42\%)\\
		\hline
		
	\end{tabular}
\end{table}

We create an aggregated categorization of bundles by adding
the category number for each bundle. We run the analysis and the
results for upgrading the bundles based on the aggregated metric is
shown in Table~\ref{table:new_bundles_aggregated}. Both the results show that the simple approach
proposed in this paper can help businesses create improved bundles
by selecting appropriate items.
\begin{table}[!h]
	\caption{Performance of New Bundles for Aggregated Metric}
	\label{table:new_bundles_aggregated}
	\begin{tabular}{l|ccc}
		\hline
		Category Shift& \multicolumn{3}{c}{Strategies}\\
		\hline
		& Replace& Add& Delete\\
		\hline
		Unpopular$\rightarrow$Popular&(63\%-71\%)&(58\%-61\%)&(38\%-40\%)\\
		Unpopular$\rightarrow$Very Popular&(38\%-41\%)&(21\%-23\%)&(13\%-16\%)\\
		Popular$\rightarrow$ Very Popular&(67\%-71\%)&(57\%-60\%)&(43\%-48\%)\\
		\hline
		
	\end{tabular}
\end{table}

We also use regression model for utility function shown in Equation~\ref{eq:coverage}. We use existing bundles to train the regression model and
predict the popularity metrics discussed in Section 3.2. The results
for the existing bundles and upgraded bundles is shown in Table~\ref{table:new_bundles_regression}.
It shows that the proposed methodology improves average performance
significantly.

\begin{table}[!h]
	\caption{Performance of New Bundles for Regression Model}
	\label{table:new_bundles_regression}
	\begin{tabular}{l|ccccc}
		\hline
		& \multicolumn{5}{c}{Metrics}\\
		\hline
		& $P_{eb}$& $P_{mb}$ & $N_{b}^0$ & $P_{b}^B$ & $D_b$\\
		\hline
		Existing Bundles&(382)&(755)&(2.75)&(226)&(0.15)\\
		Updated Bundles&(432)&(1081)&(2.02)&(345)&(0.11)\\
		\hline
		Improvement&(13\%)&(43\%)&(26\%)&(52\%)&(27\%)\\
		\hline
		
	\end{tabular}
\end{table}

\section{Conclusion}\label{sec:conclusion}
Product bundling and recommending product bundles to consumers
is attracting significant interests from stakeholders, especially researchers
and e-commerce businesses. While bundle recommendation
has footprints in extant literature, research on bundling the
items in the catalog to make these bundles is sparse. To fill this
gap, we present an intuitive holistic approach to create new bundles.
This is one of the first works on understanding the definition of
popularity of a bundle from different perspectives. We contribute
to product bundling literature by proposing new metrics and evaluating
existing bundles. For generating new bundles, we first build
content aware and context aware embeddings for games. We then
construct various machine learning models to learn how the set
of games and their features affect the popularity of a bundle. And
finally, use an iterative greedy sampling approach to create new
bundles from the existing bundles. Our results indicate that the
proposed methodology outperforms the existing bundle by 13\%-52\%
across different metrics. Top bundles generated from the proposed
methodology outperforms the existing bundles in all the five popularity
metrics.

Our results also show that we can improve the
existing bundles and shift the bundles which are unpopular to popular
and very popular category. The methodology proposed in this
paper in generic and can be implemented in different businesses.
This work has certain limitations which provides opportunities
for further research. One of the major limitations is the lack of
dataset for bundles. A research direction could be using the methodology
to check the generalize ability across different domains. Most
of the existing works consider co-purchases and co-views to make
a bundle but it is very sparse and lacks an understanding of consumer
intent. Thus, compiling datasets could be a priority work
to further this field. We use naive greedy approach to optimize the
bundles. Bundles can be improved by using more sophisticated
optimization models to sample items that could generate a set of
potentially popular bundles. An optimization model to find items
for bundling could be an interesting area of research. We focus on
creating static bundles, that is, bundles can be created and released
by the firms when a product is launched. However, we do not focus
on dynamic bundle creation based on the context of the current
session (views/purchase) as done in ecommerce websites e.g. Amazon.
This work can be extended to building dynamic bundles by
matching session intent and context.
	\bibliographystyle{ACM-Reference-Format}
	\bibliography{library, sample-base}


\begin{thebibliography}{20}


\ifx \showCODEN    \undefined \def \showCODEN     #1{\unskip}     \fi
\ifx \showDOI      \undefined \def \showDOI       #1{#1}\fi
\ifx \showISBNx    \undefined \def \showISBNx     #1{\unskip}     \fi
\ifx \showISBNxiii \undefined \def \showISBNxiii  #1{\unskip}     \fi
\ifx \showISSN     \undefined \def \showISSN      #1{\unskip}     \fi
\ifx \showLCCN     \undefined \def \showLCCN      #1{\unskip}     \fi
\ifx \shownote     \undefined \def \shownote      #1{#1}          \fi
\ifx \showarticletitle \undefined \def \showarticletitle #1{#1}   \fi
\ifx \showURL      \undefined \def \showURL       {\relax}        \fi
\providecommand\bibfield[2]{#2}
\providecommand\bibinfo[2]{#2}
\providecommand\natexlab[1]{#1}
\providecommand\showeprint[2][]{arXiv:#2}

\bibitem[Avny~Brosh et~al\mbox{.}(2022)]%
        {avny2022bruce}
\bibfield{author}{\bibinfo{person}{Tzoof Avny~Brosh}, \bibinfo{person}{Amit
  Livne}, \bibinfo{person}{Oren Sar~Shalom}, \bibinfo{person}{Bracha Shapira},
  {and} \bibinfo{person}{Mark Last}.} \bibinfo{year}{2022}\natexlab{}.
\newblock \showarticletitle{BRUCE: Bundle Recommendation Using Contextualized
  item Embeddings}. In \bibinfo{booktitle}{\emph{Proceedings of the 16th ACM
  Conference on Recommender Systems}}. \bibinfo{pages}{237--245}.
\newblock


\bibitem[Bai et~al\mbox{.}(2019)]%
        {bai2019personalized}
\bibfield{author}{\bibinfo{person}{Jinze Bai}, \bibinfo{person}{Chang Zhou},
  \bibinfo{person}{Junshuai Song}, \bibinfo{person}{Xiaoru Qu},
  \bibinfo{person}{Weiting An}, \bibinfo{person}{Zhao Li}, {and}
  \bibinfo{person}{Jun Gao}.} \bibinfo{year}{2019}\natexlab{}.
\newblock \showarticletitle{Personalized bundle list recommendation}. In
  \bibinfo{booktitle}{\emph{The World Wide Web Conference}}.
  \bibinfo{pages}{60--71}.
\newblock


\bibitem[Castells et~al\mbox{.}(2021)]%
        {castells2021novelty}
\bibfield{author}{\bibinfo{person}{Pablo Castells}, \bibinfo{person}{Neil
  Hurley}, {and} \bibinfo{person}{Saul Vargas}.}
  \bibinfo{year}{2021}\natexlab{}.
\newblock \showarticletitle{Novelty and diversity in recommender systems}.
\newblock In \bibinfo{booktitle}{\emph{Recommender systems handbook}}.
  \bibinfo{publisher}{Springer}, \bibinfo{pages}{603--646}.
\newblock


\bibitem[Cer et~al\mbox{.}(2018)]%
        {cer2018universal}
\bibfield{author}{\bibinfo{person}{Daniel Cer}, \bibinfo{person}{Yinfei Yang},
  \bibinfo{person}{Sheng-yi Kong}, \bibinfo{person}{Nan Hua},
  \bibinfo{person}{Nicole Limtiaco}, \bibinfo{person}{Rhomni~St John},
  \bibinfo{person}{Noah Constant}, \bibinfo{person}{Mario Guajardo-Cespedes},
  \bibinfo{person}{Steve Yuan}, \bibinfo{person}{Chris Tar}, {et~al\mbox{.}}}
  \bibinfo{year}{2018}\natexlab{}.
\newblock \showarticletitle{Universal sentence encoder}.
\newblock \bibinfo{journal}{\emph{arXiv preprint arXiv:1803.11175}}
  (\bibinfo{year}{2018}).
\newblock


\bibitem[Church(2017)]%
        {church2017word2vec}
\bibfield{author}{\bibinfo{person}{Kenneth~Ward Church}.}
  \bibinfo{year}{2017}\natexlab{}.
\newblock \showarticletitle{Word2Vec}.
\newblock \bibinfo{journal}{\emph{Natural Language Engineering}}
  \bibinfo{volume}{23}, \bibinfo{number}{1} (\bibinfo{year}{2017}),
  \bibinfo{pages}{155--162}.
\newblock


\bibitem[Gholami et~al\mbox{.}(2022)]%
        {gholami2022parsrec}
\bibfield{author}{\bibinfo{person}{Ehsan Gholami}, \bibinfo{person}{Mohammad
  Motamedi}, {and} \bibinfo{person}{Ashwin Aravindakshan}.}
  \bibinfo{year}{2022}\natexlab{}.
\newblock \showarticletitle{PARSRec: Explainable personalized attention-fused
  recurrent sequential recommendation using session partial actions}. In
  \bibinfo{booktitle}{\emph{Proceedings of the 28th ACM SIGKDD Conference on
  Knowledge Discovery and Data Mining}}. \bibinfo{pages}{454--464}.
\newblock


\bibitem[Li et~al\mbox{.}(2017)]%
        {li2017neural}
\bibfield{author}{\bibinfo{person}{Jing Li}, \bibinfo{person}{Pengjie Ren},
  \bibinfo{person}{Zhumin Chen}, \bibinfo{person}{Zhaochun Ren},
  \bibinfo{person}{Tao Lian}, {and} \bibinfo{person}{Jun Ma}.}
  \bibinfo{year}{2017}\natexlab{}.
\newblock \showarticletitle{Neural attentive session-based recommendation}. In
  \bibinfo{booktitle}{\emph{Proceedings of the 2017 ACM on Conference on
  Information and Knowledge Management}}. \bibinfo{pages}{1419--1428}.
\newblock


\bibitem[Mikolov et~al\mbox{.}(2018)]%
        {mikolov2018advances}
\bibfield{author}{\bibinfo{person}{Tomas Mikolov}, \bibinfo{person}{Edouard
  Grave}, \bibinfo{person}{Piotr Bojanowski}, \bibinfo{person}{Christian
  Puhrsch}, {and} \bibinfo{person}{Armand Joulin}.}
  \bibinfo{year}{2018}\natexlab{}.
\newblock \showarticletitle{Advances in Pre-Training Distributed Word
  Representations}. In \bibinfo{booktitle}{\emph{Proceedings of the
  International Conference on Language Resources and Evaluation (LREC 2018)}}.
\newblock


\bibitem[Nayak et~al\mbox{.}(2023)]%
        {nayak2023news}
\bibfield{author}{\bibinfo{person}{Ashutosh Nayak}, \bibinfo{person}{Mayur
  Garg}, {and} \bibinfo{person}{Rajasekhara~Reddy Duvvuru~Muni}.}
  \bibinfo{year}{2023}\natexlab{}.
\newblock \showarticletitle{News Popularity Beyond the Click-Through-Rate for
  Personalized Recommendations}. In \bibinfo{booktitle}{\emph{Proceedings of
  the 46th International ACM SIGIR Conference on Research and Development in
  Information Retrieval}}. \bibinfo{pages}{1396--1405}.
\newblock


\bibitem[Pathak et~al\mbox{.}(2017)]%
        {pathak2017generating}
\bibfield{author}{\bibinfo{person}{Apurva Pathak}, \bibinfo{person}{Kshitiz
  Gupta}, {and} \bibinfo{person}{Julian McAuley}.}
  \bibinfo{year}{2017}\natexlab{}.
\newblock \showarticletitle{Generating and personalizing bundle recommendations
  on steam}. In \bibinfo{booktitle}{\emph{Proceedings of the 40th International
  ACM SIGIR Conference on Research and Development in Information Retrieval}}.
  \bibinfo{pages}{1073--1076}.
\newblock


\bibitem[Puthiya~Parambath et~al\mbox{.}(2016)]%
        {puthiya2016coverage}
\bibfield{author}{\bibinfo{person}{Shameem~A Puthiya~Parambath},
  \bibinfo{person}{Nicolas Usunier}, {and} \bibinfo{person}{Yves Grandvalet}.}
  \bibinfo{year}{2016}\natexlab{}.
\newblock \showarticletitle{A coverage-based approach to recommendation
  diversity on similarity graph}. In \bibinfo{booktitle}{\emph{Proceedings of
  the 10th ACM Conference on Recommender Systems}}. \bibinfo{pages}{15--22}.
\newblock


\bibitem[Reimers(2019)]%
        {reimers2019sentence}
\bibfield{author}{\bibinfo{person}{N Reimers}.}
  \bibinfo{year}{2019}\natexlab{}.
\newblock \showarticletitle{Sentence-BERT: Sentence Embeddings using Siamese
  BERT-Networks}.
\newblock \bibinfo{journal}{\emph{arXiv preprint arXiv:1908.10084}}
  (\bibinfo{year}{2019}).
\newblock


\bibitem[Seo et~al\mbox{.}(2022)]%
        {seo2022ta}
\bibfield{author}{\bibinfo{person}{Jaejin Seo}, \bibinfo{person}{Sangwon Lee},
  \bibinfo{person}{Ling Liu}, {and} \bibinfo{person}{Wonik Choi}.}
  \bibinfo{year}{2022}\natexlab{}.
\newblock \showarticletitle{TA-SBERT: token attention sentence-BERT for
  improving sentence representation}.
\newblock \bibinfo{journal}{\emph{IEEE Access}}  \bibinfo{volume}{10}
  (\bibinfo{year}{2022}), \bibinfo{pages}{39119--39128}.
\newblock


\bibitem[Song et~al\mbox{.}(2020)]%
        {song2020towards}
\bibfield{author}{\bibinfo{person}{Qingquan Song}, \bibinfo{person}{Dehua
  Cheng}, \bibinfo{person}{Hanning Zhou}, \bibinfo{person}{Jiyan Yang},
  \bibinfo{person}{Yuandong Tian}, {and} \bibinfo{person}{Xia Hu}.}
  \bibinfo{year}{2020}\natexlab{}.
\newblock \showarticletitle{Towards automated neural interaction discovery for
  click-through rate prediction}. In \bibinfo{booktitle}{\emph{Proceedings of
  the 26th ACM SIGKDD International Conference on Knowledge Discovery \& Data
  Mining}}. \bibinfo{pages}{945--955}.
\newblock


\bibitem[Sun et~al\mbox{.}(2022)]%
        {sun2022revisiting}
\bibfield{author}{\bibinfo{person}{Zhu Sun}, \bibinfo{person}{Jie Yang},
  \bibinfo{person}{Kaidong Feng}, \bibinfo{person}{Hui Fang},
  \bibinfo{person}{Xinghua Qu}, {and} \bibinfo{person}{Yew~Soon Ong}.}
  \bibinfo{year}{2022}\natexlab{}.
\newblock \showarticletitle{Revisiting Bundle Recommendation: Datasets, Tasks,
  Challenges and Opportunities for Intent-aware Product Bundling}. In
  \bibinfo{booktitle}{\emph{Proceedings of the 45th International ACM SIGIR
  Conference on Research and Development in Information Retrieval}}.
  \bibinfo{pages}{2900--2911}.
\newblock


\bibitem[Turgut et~al\mbox{.}(2023)]%
        {turgut2023prod2vec}
\bibfield{author}{\bibinfo{person}{Hacer Turgut}, \bibinfo{person}{Tan~Doruk
  Yetki}, \bibinfo{person}{{\"O}m{\"u}r Bali}, {and}
  \bibinfo{person}{Tayfun~Arda Y{\"u}cel}.} \bibinfo{year}{2023}\natexlab{}.
\newblock \showarticletitle{Prod2Vec-Var: A Session Based Recommendation System
  with Enhanced Diversity}. In \bibinfo{booktitle}{\emph{Proceedings of the
  32nd ACM International Conference on Information and Knowledge Management}}.
  \bibinfo{pages}{5253--5254}.
\newblock


\bibitem[Vasile et~al\mbox{.}(2016)]%
        {vasile2016meta}
\bibfield{author}{\bibinfo{person}{Flavian Vasile}, \bibinfo{person}{Elena
  Smirnova}, {and} \bibinfo{person}{Alexis Conneau}.}
  \bibinfo{year}{2016}\natexlab{}.
\newblock \showarticletitle{Meta-prod2vec: Product embeddings using
  side-information for recommendation}. In
  \bibinfo{booktitle}{\emph{Proceedings of the 10th ACM conference on
  recommender systems}}. \bibinfo{pages}{225--232}.
\newblock


\bibitem[Vaswani et~al\mbox{.}(2017)]%
        {vaswani2017attention}
\bibfield{author}{\bibinfo{person}{Ashish Vaswani}, \bibinfo{person}{Noam
  Shazeer}, \bibinfo{person}{Niki Parmar}, \bibinfo{person}{Jakob Uszkoreit},
  \bibinfo{person}{Llion Jones}, \bibinfo{person}{Aidan~N Gomez},
  \bibinfo{person}{{\L}ukasz Kaiser}, {and} \bibinfo{person}{Illia
  Polosukhin}.} \bibinfo{year}{2017}\natexlab{}.
\newblock \showarticletitle{Attention is all you need}.
\newblock \bibinfo{journal}{\emph{Advances in neural information processing
  systems}}  \bibinfo{volume}{30} (\bibinfo{year}{2017}).
\newblock


\bibitem[Wang and Kuo(2020)]%
        {wang2020sbert}
\bibfield{author}{\bibinfo{person}{Bin Wang} {and} \bibinfo{person}{C-C~Jay
  Kuo}.} \bibinfo{year}{2020}\natexlab{}.
\newblock \showarticletitle{Sbert-wk: A sentence embedding method by dissecting
  bert-based word models}.
\newblock \bibinfo{journal}{\emph{IEEE/ACM Transactions on Audio, Speech, and
  Language Processing}}  \bibinfo{volume}{28} (\bibinfo{year}{2020}),
  \bibinfo{pages}{2146--2157}.
\newblock


\bibitem[Yadav and Monroe(1993)]%
        {yadav1993buyers}
\bibfield{author}{\bibinfo{person}{Manjit~S Yadav} {and}
  \bibinfo{person}{Kent~B Monroe}.} \bibinfo{year}{1993}\natexlab{}.
\newblock \showarticletitle{How buyers perceive savings in a bundle price: An
  examination of a bundle's transaction value}.
\newblock \bibinfo{journal}{\emph{Journal of Marketing Research}}
  \bibinfo{volume}{30}, \bibinfo{number}{3} (\bibinfo{year}{1993}),
  \bibinfo{pages}{350--358}.
\newblock


\end{thebibliography}
	
\end{document}